\documentclass[twocolumn]{aastex63}

\usepackage{graphicx}	
\usepackage{amsmath}	
\usepackage{amssymb}	
\usepackage{gensymb}
\shorttitle{The life cycle of magnetars}
\shortauthors{Tushar Mondal}

\begin{document}

\title{The life cycle of magnetars: a novel approach to estimate their ages}

\correspondingauthor{Tushar Mondal}
\email{mtushar@iisc.ac.in, tusharmondal770@gmail.com}

\author[0000-0001-8174-2011]{Tushar Mondal}
\affiliation{Department of Physics, Indian Institute of Science, Bangalore 560012, India}



\begin{abstract}

Anomalous X-ray pulsars and soft gamma repeaters are slowly rotating, young, and isolated neutron stars exhibiting sporadic outbursts and high X-ray quiescent luminosities. They are believed to be powered by ultrastrong magnetic fields, $B\sim10^{14}-10^{15}$ G, associated with `magnetars'. In the peculiar case of SGR 0418+5729, timing parameters imply a dipolar $B$-field of $6.1\times10^{12}$ G. This discovery has challenged the traditional picture of magnetars in terms of $B$-field strengths, evolutionary stages, and ages. Here we provide a novel approach to estimate a magnetar's age by considering the self-consistent time evolution of a plasma-filled oblique pulsar with the state-of-the-art magnetospheric particle acceleration gaps. The rotational period of magnetars increases over time due to angular momentum extraction by gravitational-wave radiations, magnetic dipole radiations, and particle winds. These torques also change the obliquity angle between the magnetic and rotation axes. For SGR 0418+5729, we obtain a dipolar $B$-field of $1.0\times10^{14}$ G, and a realistic age of $\sim18$ kyr, consistent within the magnetar paradigm.

\end{abstract}



\section{Introduction} \label{sec:intro}

Magnetized, rotating, isolated neutron stars often exhibit pulsation in the radio and X-ray bands, hence the name pulsars. The rotational period of pulsars increases over time as their rotational energy is lost due to magnetic dipole radiation. Based on an orthogonal vacuum rotator model, the characteristic surface dipole magnetic field $(B_{c})$ of a pulsar can be estimated from its observed spin period, $P$, and spin-down rate, $\dot{P}$, as given by $B_{c}=(3Ic^{3}P\dot{P}/8\pi^{2}R^{6})^{1/2}\simeq 3.2\times 10^{19}(P\dot{P})^{1/2}$ G, where $c$ is the speed of light, and we assume the neutron star radius $R=10^{6}$ cm, and the moment of inertia $I=10^{45}$ g cm$^{2}$. One can also obtain the characteristic age of a pulsar based on that magnetic dipole spin-down model, as given by $\tau_{c}=P/2\dot{P}$.

The expression for $B_{c}$ was originally developed to estimate the magnetic fields of ordinary pulsars, usually up to $\sim 10^{12}$ G. However, it has been traditionally used also for magnetars, in which the derived values of $B_{c}$ appear to be as high as $\sim 10^{14}-10^{15}$ G\footnote{\url{www.physics.mcgill.ca/~pulsar/magnetar/main.html}}. Soft gamma repeaters (SGRs) and anomalous X-ray pulsars (AXPs) are two observational manifestations of magnetars and have been studied extensively with every modern X-ray telescope. They are a special class of X-ray pulsars with observational properties that differ from the more common accretion-powered X-ray pulsars and the rotation-powered radio pulsars. AXPs are distinguished from X-ray binaries by their soft X-ray spectrum, lack of evidence of their binary companions, narrow spin period distribution, and fast spin-down rate. On the other hand, SGRs were discovered by detecting the short bursts in the hard X-ray/soft gamma-ray range and are now considered a separate class of objects from the classical gamma-ray bursts. Both SGRs and AXPs share a number of interesting properties \citep{2008A&ARv..15..225M}: rotational periods in the range of $P\sim 2-12$ s, a narrow range compared to that for the ordinary pulsars;
spin-down rates of $\dot{P}\sim 10^{-13}-10^{-10}$ s/s, larger than that for the ordinary pulsars;
persistent X-ray luminosity in the range of $10^{33}-10^{36}$ ergs/s, generally much higher than the rate of rotational energy losses; and sporadic bursting activity.

Within the magnetar paradigm, the rapid decay of the intense magnetic fields powers the persistent emission, while the observed bursts are attributed either to crustal quakes produced by magnetic stresses \citep{1995MNRAS.275..255T} or violent magnetic reconnections \citep{2003MNRAS.346..540L}. The magnetar-like activity is generally associated with ultrastrong magnetic fields, typically higher than the electron quantum field, $B_{Q}=m_{e}^{2}c^{3}/e\hbar \simeq 4.4\times 10^{13}$ G. Theoretical work indicates that the quantum electrodynamic effects (photon splitting, for example) become important, and the pulsar radio emission is suppressed when the neutron star magnetic field $B$ reaches or exceeds  $B_{Q}$ \citep{1998ApJ...507L..55B}. The super-strong magnetic fields, in principle up to $10^{17}\times$ (3 ms$/P_{i})$ G, can be produced via an active dynamo in a differentially rotating proto-neutron star with an initial spin period $(P_{i})$ shorter than the convective overturn time of $\sim 3-10$ ms  \citep{1993ApJ...408..194T}. 

This traditional picture of magnetars is challenged by the discovery of unusual low magnetic field magnetar SGR 0418+5729 \citep{2010Sci...330..944R}. The measured $P=9.08$ s and $\dot{P}=4.0\times 10^{-15}$ s/s translate into a dipole magnetic field $B_{c}=6.1\times 10^{12}$ G, well in the range of ordinary radio pulsars. However, SGR 0418+5729 exhibits all other magnetar-like activities: $(a)$ emission of sporadic outbursts, $(b)$ enhanced persistent flux, and $(c)$ an X-ray spectrum characterized by a blackbody plus non-thermal power-law component, which softened during the outburst decay. Interestingly, the X-ray spectral analysis strongly disfavoured the neutron star atmosphere models with moderate surface magnetic field strengths ($10^{12}-10^{13}$ G), and the best-fitting value of magnetic field strength at the surface was found as $1.0\times 10^{14}$ G \citep{2011MNRAS.418.2773G}. Also, a variable absorption feature has been detected in the X-ray spectrum of SGR 0418+5729. The line was interpreted as a proton-cyclotron resonant scattering feature, and the line energy implied a surface magnetic field of $\gtrsim 2\times 10^{14}$ G \citep{2013Natur.500..312T}. Hence, there is a large discrepancy between the magnetic field strength inferred from the spin-down feature and the magnetic field strength inferred from the spectral analysis. A similar discrepancy has been found in the second low-field magnetar SWIFT J1822.3-1606. Its timing properties ($P=8.4377$ s, and $\dot{P}=2.1\times 10^{-14}$ s/s) indicate a dipole magnetic field $B_{c}=1.4\times 10^{13}$ G, whereas the presence of variable cyclotron feature implies much stronger ($\gtrsim10^{14}$ G) surface magnetic fields \citep{2016MNRAS.456.4145R}.

Another challenging issue is the actual age measurement of magnetars. This is very crucial to estimate the formation rate of Galactic magnetars. It remains mysterious over decades to understand the birth site for such a distinct population of neutron stars. The magnetar birth rate may unveil their true progenitor. Key information on SGRs/AXPs can be obtained by their association with supernova remnants (SNRs). 
Such an association provides the opportunity to estimate the magnetar's age, distance, and space velocity. The genuine associations between radio pulsars and SNRs are judged based on the agreement in pulsar's $\tau_{c}$, and the transverse velocity. However, for magnetars, $\tau_{c}$ is not a reliable indicator for their true ages \citep{2015PASJ...67....9N}. $\tau_{c}$ reflects the true age only if the neutron star's spin-down is entirely due to magnetic dipole radiation without considering decay of the magnetic fields. But, for magnetars, magnetic field decay is very fundamental, and significant additional torques are expected due to relativistic particle winds and gravitational wave (GW) emissions. Hence, there is no expectation to estimate the true ages of SGRs and AXPs from their spin parameters only. The range of transverse velocity is also not clear for SGRs and AXPs. Therefore we need to rely only on the positional coincidence on the sky to identify SNR associations \citep{2001ApJ...559..963G}. There remain some possibilities that an AXP/SGR and an adjacent SNR are physically unrelated \citep{2012ApJ...761...76T}. The realistic age measurement of AXPs/SGRs can confirm such genuine associations independently.

In this Letter, we resolve both the issues: $(a)$ the discrepancy in the surface magnetic field measurement between the timing and spectral analysis, and $(b)$ the actual age measurement of magnetars independently.

\section{Magnetar spin-down} \label{sec:spin-down}

In pulsar theory, the magnetic dipole approximation is often assumed to estimate $B_{c}$ and $\tau_{c}$ for both ordinary radio pulsars and magnetars. Such magnetic dipole approximation is just a pedagogical model, in which an orthogonal rotating dipole is considered in the vacuum. These vacuum magnetosphere solutions can not produce any pulsar-like emission. In general, a real pulsar should be an oblique rotator surrounded by a plasma-filled magnetosphere. In this plasma, the particle acceleration is very crucial to generate pulsar radio and high-energy emission. Also, the plasma effects in the pulsar magnetosphere can substantially modify the spin-down torques. It has been confirmed numerically through force-free electrodynamics for aligned rotators \citep{2005PhRvL..94b1101G} and for oblique rotators \citep{2006ApJ...648L..51S}, resistive magnetohydrodynamics \citep{2012ApJ...746...60L}, as well as particle-in-cell \citep{2015ApJ...801L..19P} simulations.

Besides strong surface dipole magnetic fields, magnetars are generally considered to harbor even stronger internal toroidal fields \citep{2002ApJ...574..332T,2011ApJ...741..123P}. Observationally, the slow phase modulation in the hard X-ray emission from 4U 0142+61 as free precession of the neutron star \citep{2014PhRvL.112q1102M}, the powerful giant flare from SGR 1806-20 \citep{2005ApJ...634L.165S} all indicate that the internal toroidal fields are at least one order of magnitude stronger than the external dipole field of a magnetar, reaching $\sim 10^{16}$ G or higher. Throughout, we consider the ratio of toroidal to poloidal components of magnetic fields is 20. However, we find that this ratio with a value $\lesssim30$ plays hardly any role in the spin-down evolution of magnetar's life cycle.

The anisotropic pressure from the magnetic fields deforms the star into a prolate or an oblate shape depending on the toroidal, or poloidal fields dominated configurations, respectively. Such deformations are taken care of through the ellipticity parameter, $\epsilon$, based on the magnetically deformed neutron star models \citep{2008MNRAS.385..531H,2011MNRAS.417.2288M}, as given by
\begin{multline*}
\epsilon = 6.262\times 10^{-6}\left(\frac{B}{5\times 10^{14}\ \text{G}}\right)^{2} \left(\frac{M}{1.4M_{\odot}}\right)^{-2}\times \\ \left(\frac{R}{10\ \text{km}}\right)^{4}  \left(1-\frac{0.385}{\Lambda}\right),
\end{multline*}
where $B$ is the surface dipole magnetic field, $\Lambda$ is the ratio of poloidal to total magnetic field energy.
As long as this magnetically induced deformation is asymmetric with respect to the rotation axis, the star's spin causes a time-varying mass quadrupole moment, which leads to the GW radiations. The rotational period of both isolated pulsars and magnetars increases over time due to angular momentum extraction by gravitational and electromagnetic torques. The electromagnetic torque consists of magnetic dipole radiations and particle winds. These torques also change the obliquity angle, $\chi$, between the rotation and magnetic axes. For an oblique rotator, the evolution equations (spin-down and alignment) will be a combination of GW radiations \citep{1970Natur.228..655C,2020ApJ...896...69K} and two electromagnetic energy-loss terms \citep{2006ApJ...643.1139C,2012ApJ...746...60L}, as given by
\begin{multline}
	I\frac{d\Omega}{dt}=-T_{\text{G}}\epsilon^{2}\Omega^{5}\sin^{2}\chi(1+15\sin^{2}\chi)-T_{\text{E}}B^{2}\Omega^{3}\times \\
	\begin{cases}
		\left[1.1\sin^{2}\chi+0.9\left(1-\frac{\Omega_{\text{death}}}{\Omega}\right)\right],\ \text{if}\ \Omega>\Omega_{\text{death}} \\
		1.1\sin^{2}\chi, \hspace{3.25cm} \text{if}\ \Omega\leq\Omega_{\text{death}}
	\end{cases}, \label{eq: spin-down}
\end{multline}
\begin{equation}
	I\Omega\frac{d\chi}{dt}=-6T_{\text{G}}\epsilon^{2}\Omega^{5}\sin^{3}\chi\cos\chi-1.1T_{\text{E}}B^{2}\Omega^{3}\sin\chi\cos\chi, \label{eq: allignment}
\end{equation}
where $\Omega$ is the angular velocity. Here, the terms associated with $T_{\text{G}}=2GI^{2}/(5c^{5})$ and $T_{\text{E}}=R^{6}/(4c^{3})$ are the torques from GW radiation and electromagnetic mechanisms respectively. The electromagnetic spin-down torque, as in Eq. \ref{eq: spin-down}, is a combination of two mechanisms: one proportional to $\sin^{2}\chi$ is due to magnetic dipole radiation, and the remaining term is due to particle wind. $\Omega_{\text{death}}=2\pi/P_{\text{death}}$ is the angular velocity to describe the pulsar `death', i.e., the stopping of any pulsar-like emission, and the death period can be expressed as
\begin{equation*}
	P_{\text{death}}=2.84\left(\frac{V_{\text{gap}}}{10^{13}\ \text{V}}\right)^{-1/2}\left(\frac{B}{10^{13}\ \text{G}}\right)^{1/2} \ \text{s} , \label{eq: death period}
\end{equation*}
where $V_{\text{gap}}$ is the particle acceleration gap potential developed along open magnetic field lines. Typically, $V_{\text{gap}}=10^{13}$ V is used in the model of particle acceleration of pulsars \citep{2006ApJ...643.1139C,2012ApJ...757L..10T}. As the neutron star slows down with time, $\Omega$ may drop below $\Omega_{\text{death}}$. Pulsars with $\Omega\leq\Omega_{\text{death}}$ can not maintain the required gap potential for particle acceleration, and hence, the torque associated with particle wind will not operate. Beyond the death, the misaligned neutron star will continue to spin down through GW radiations and magnetic dipole radiations without generating any pulse.
At $P=P_{\text{death}}$, we have (from Eq.~\ref{eq: spin-down}),
\begin{multline}
I\frac{d\Omega_{\text{death}}}{dt}=-T_{\text{G}}\epsilon^{2}\Omega_{\text{death}}^{5}\sin^{2}\chi(1+15\sin^{2}\chi)- \\ T_{\text{E}}B^{2}\Omega_{\text{death}}^{3} (1.1\sin^{2}\chi). \label{eq: death line}
\end{multline}
Hence, the death line $\dot{P}(P_{\text{death}})$ has an angular dependence.

For magnetars, rotational energy is not the ultimate source of energy. Rapid decay of ultrastrong magnetic fields is the main energy reservoir to fuel the SGR/AXP activity. The three main avenues of magnetic field decay in isolated neutron stars are Ohmic dissipation, ambipolar diffusion, and Hall cascade \citep{1992ApJ...395..250G}. Ambipolar diffusion only operates in the neutron star core and plays a very lesser role in the active lifetimes of magnetars. It has been shown that the magnetar activity is controlled by a decaying magnetic field outside the core and governed by Hall drift and Ohmic dissipation at the stellar crust \citep{2000ApJ...529L..29C,2011MNRAS.413.2021G}. Recent advances in coupled magneto-thermal evolution of isolated neutron stars allow one to self-consistently account for the dipole magnetic field evolution, as given by \citep{2008A&A...486..255A}
\begin{equation}
	\frac{dB}{dt}=-\frac{B}{\tau_{\text{Ohm}}}-\frac{1}{B_{i}}\frac{B^{2}}{\tau_{\text{Hall}}}, \label{eq: field_decay}
\end{equation}
where $B_{i}$ is the initial magnetic field strength, and $\tau_{\text{Ohm}}$ and $\tau_{\text{Hall}}$ are the characteristic timescales of Ohmic and Hall decay, respectively. 
Note that we do not include the effect of the Hall attractor stage \citep{2014PhRvL.112q1101G}, an equilibrium state, in this phenomenological description of the magnetic field decay. This assumption is viable and has a lesser role in the age measurement since the stage of the Hall attractor starts after a few Hall time-scales \citep{2014MNRAS.438.1618G, 2015AN....336..831I} and by the time about 90\% of the magnetic energy has already decayed \citep{2016PNAS..113.3944G}.
The ratio of the Ohmic to Hall timescale is equivalent to the product of electron cyclotron frequency and electron collision time, and it strongly depends on the temperature of the crust. For magnetar range surface magnetic fields, this ratio turns out to be $\sim 1-10 B_{i,13}$, where $B_{i,13}$ is $B_{i}$ in units of $10^{13}$ G \citep{2004ApJ...609..999C,2007A&A...470..303P}. Hence, the Hall drift affects drastically the very early evolution of strongly magnetized neutron stars. For magnetars, we have taken $\tau_{\text{Ohm}}=10^{6}$ yr and $\tau_{\text{Hall}}=2\times 10^{3}\ \text{yr}/B_{i,15}$, i.e., $\tau_{\text{Ohm}}/\tau_{\text{Hall}}=5 B_{i,13}$ throughout in our computation \citep{2008ApJ...673L.167A}.

\section{Methods} \label{sec:methods}

One needs to solve the set of three coupled differential Eqs. \ref{eq: spin-down}, \ref{eq: allignment}, and \ref{eq: field_decay} simultaneously using appropriate initial conditions to obtain the solutions for three dynamical variables: $\Omega$, $\chi$, and $B$, as a function of time $t$. The present value of any variable is denoted by subscript zero. From observations (timing properties), we have the present values of angular velocity $\Omega_{0}$, and its time derivative $(d\Omega/dt)_{0}$. With a chosen $\chi_{0}$, where $0\leq  \sin^{2}\chi_{0} \leq1$, we can obtain a specific value of $B_{0}$ satisfying Eq. \ref{eq: spin-down}. Assigning these $\Omega_{0}$, $\chi_{0}$, and $B_{0}$, the integration of Eqs. \ref{eq: spin-down}, \ref{eq: allignment}, and \ref{eq: field_decay} is carried backward in time starting from $t=0$. The termination point $-t_{z}$ is the value of $t$ in the limit of $\chi \rightarrow \pi/2$. The reasons are as follows. Newborn magnetars experience an evolution where internal viscous damping of precession drives $\chi \rightarrow \pi/2$ shortly after birth, typically within a minute \citep{2002PhRvD..66h4025C, 2020MNRAS.494.4838L}. In general, the rotational energy of a prolate ellipsoid (i.e., one with a dominantly toroidal magnetic field) is minimized when the symmetry axis is orthogonal to the spin axis. Previously, many authors have applied this $\chi \rightarrow \pi/2$ scenario to newborn magnetars in the context of optimal GW emission \citep{2002PhRvD..66h4025C, 2005ApJ...634L.165S}. Hence, for a given $\chi_{0}$, we have a specific value of the termination point $-t_{z}$, which is nothing but the age of the corresponding magnetar. In another way, if we know the actual age of a magnetar, we can specify the exact value of the obliquity angle $\chi_{0}$. In case we do not have any information of both the actual age and the obliquity angle, we can address the bound on age based on the bound on $0\leq \sin^{2}\chi_{0} \leq1$. 

\section{Results and discussions} \label{sec:results}

\begin{figure}
	\center
	\includegraphics[width=\columnwidth]{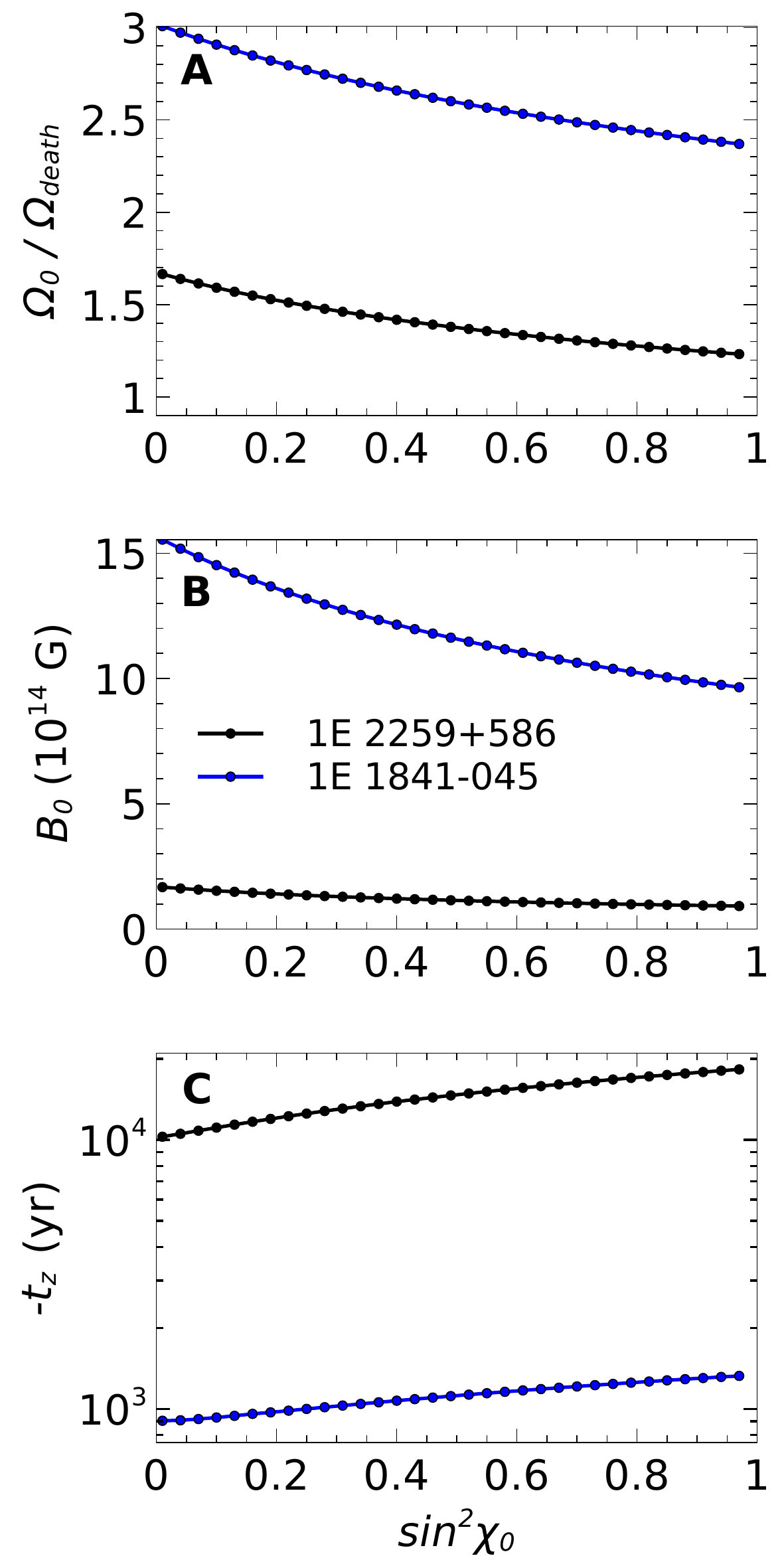}\caption{Variations of (A) the present value of angular velocity over the corresponding `death', (B) the present value surface dipole $B$-field, and (C) the model age, as a function of the present value of the obliquity angle.
	} 
	\label{fig:high_field_magnetar}
\end{figure}
First, we consider some specific sources that follow the traditional picture of magnetars and the associated SNRs of which are known. AXP 1E 2259+586 is associated with the SNR CTB 109 with age $\approx 14$ kyr \citep{2013A&A...552A..45S,2015PASJ...67....9N}. However, the characteristic age of 1E 2259+586 is $\tau_{c}=230$ kyr. Another younger source, 1E 1841-045, is associated with the SNR Kes 73. The age of SNR Kes 73 is only $\approx 1300$ yr \citep{2006MNRAS.370L..14V,2014ApJ...781...41K}, whereas $\tau_{c}$ for 1E 1841-045 is 4600 yr. Hence, the characteristic age is an overestimate of the actual age of magnetars. 
Since there are some uncertainties in the SNRs' age measurement observationally, we have not taken this parameter as an input in our model. 
Fig.~\ref{fig:high_field_magnetar}(C) indicates that the present value of the obliquity angle $\chi_{0}$ and the age of the magnears are strongly correlated. It also provides the bound on age for both the magnetars and their associated SNRs independently based on the bound on $0\leq \sin^2\chi \leq 1$. Such bound is $1.03-1.83\times 10^{4}$ yr for 1E 2259+586, and $0.90-1.33\times 10^{3}$ yr for 1E 1841-045. The corresponding dipole $B$-fields are given in Fig.~\ref{fig:high_field_magnetar}(B), and we have $B_{0}=0.91-1.67\times 10^{14}$ G for 1E 2259+586, whereas $B_{0}=9.65-15.54\times 10^{14}$ G for 1E 1841-045.
Fig.~\ref{fig:high_field_magnetar}(A) explains that both the sources are well above their corresponding `death', i.e., $\Omega_0>\Omega_{\text{death}}$. This is necessary to show any pulsar-like activities. However, compared to 1E 1841-045, 1E 2259+586 is closer to its `death'.

\begin{figure}
	\center
	\includegraphics[width=\columnwidth]{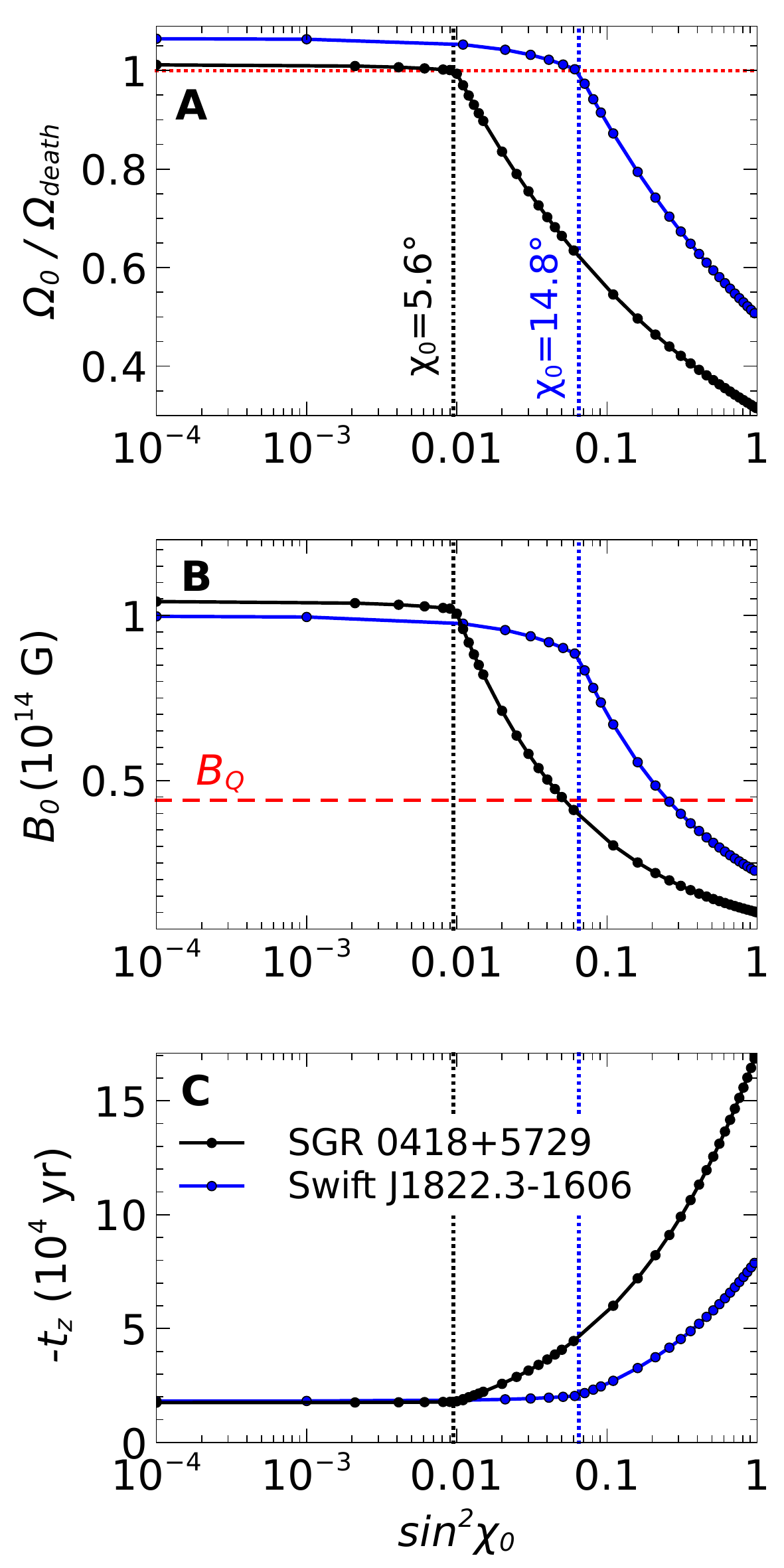}\caption{Variations of (A) the present value of angular velocity over the corresponding `death', (B) the present value surface dipole $B$-field, and (C) the model age, as a function of the present value of the obliquity angle. The red dashed line represents the electron quantum field limit. The vertical dotted lines (black: SGR 0418+5729, blue: Swift J1822.3-1606) correspond to the transition value of $\chi_{0}$ below which the sources remain above their corresponding `death', i.e., $\Omega_0>\Omega_{\text{death}}$.
	} 
	\label{fig:low_field_magnetar}
\end{figure}
In Fig.~\ref{fig:low_field_magnetar}, we consider the so-called low-magnetic field magnetars: SGR 0418+5729 and Swift J1822.3-1606. Fig.~\ref{fig:low_field_magnetar}(A) indicates that SGR 0418+5729 lies far below its `death' for obliquity angle $\chi_{0}\ge5.6\degree$. The corresponding $\chi_{0}$ for Swift J1822.3-1606 is $14.8\degree$, above which the source remains below its corresponding `death'. The `death' information provides the proper bound on $\chi_{0}$ since the rotation-powered magnetospheric activity stops below the `death'. However, both the sources appear with pulsar-like high-energy emissions, and hence, they should lie above their corresponding `death'. So, the accepted values of $\chi_{0}$ for SGR 0418+5729 and Swift J1822.3-1606 are  $\chi_{0}<5.6\degree$ and $\chi_{0}<14.8\degree$, respectively. An earlier study also pointed out the possibility of such small obliquity angles \citep{2012ApJ...757L..10T}. Using these $\chi_{0}$, we find that the surface dipole magnetic fields are $B_{0}\sim (1.01-1.04)\times 10^{14}$ G for SGR 0418+5729, and $B_{0}\sim (8.84-9.97)\times 10^{13}$ G for Swift J1822.3-1606, as given in Fig.~\ref{fig:low_field_magnetar}(B). The measured values of $B_{0}$ are well above the quantum critical field $B_{Q}$, and hence, they are actually normal magnetar instead of low magnetic field magnetars. For comparison, one can see the surface dipole $B$-fields for both the sources at $\chi_{0}=90\degree$ are well below the $B_{Q}$. Based on this, the traditional theory of magnetars has been challenged for over a decade. The corresponding age is prescribed in Fig.~\ref{fig:low_field_magnetar}(C). We find the true age of SGR 0418+5729 is $17.48-18.05$ kyr using $\chi_{0}<5.6\degree$, whereas the true age for Swift J1822.3-1606 is $18.2-20.4$ kyr using $\chi_{0}<14.8\degree$. Hence, both the surface dipole $B$-fields and the actual age are consistent with the traditional picture of magnetars.

\begin{figure}
	\center
	\includegraphics[width=\columnwidth]{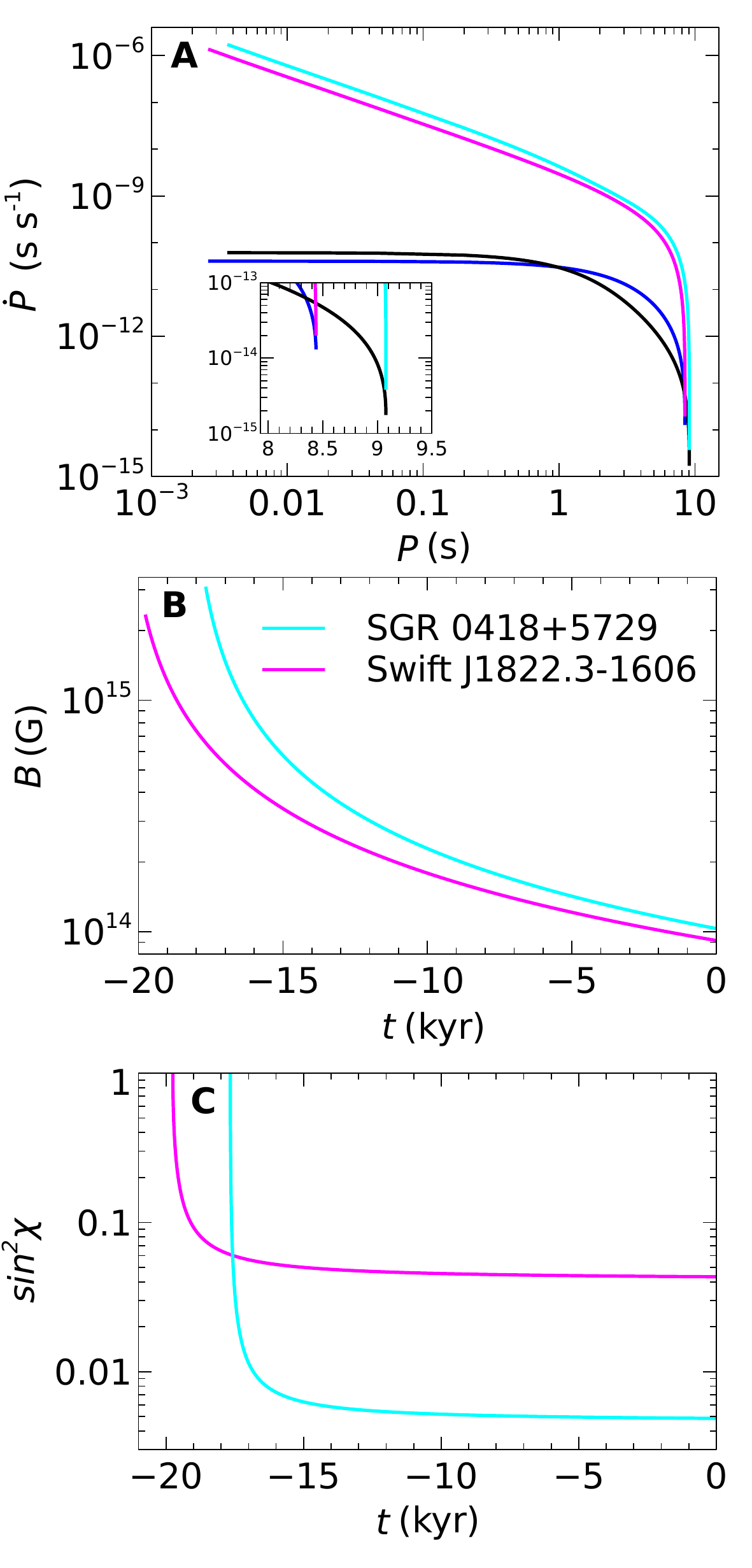}\caption{(A) Evolutionary tracks of various SGRs/AXPs in the $P-\dot{P}$ diagram along with their death line (black: SGR 0418+5729, blue: Swift J1822.3-1606). Evolution of (B) surface dipole $B$-fields, and (C) the obliquity angle of the corresponding SGRs/AXPs. Here $t=0$ corresponds to the present position, and negative termination time corresponds to the real age of the corresponding magnetar.
	} 
	\label{fig:evolution}
\end{figure}

So far, we have assigned the present value of $\chi$ and the corresponding dipole $B$-fields. Using these, the complete evolutions are shown in Fig.~\ref{fig:evolution}. Here we consider the typical values of $\chi_{0}$ are $4\degree$ and $12\degree$ for SGR 0418+5729 and Swift J1822.3-1606, respectively (see Fig.~\ref{fig:low_field_magnetar}). Fig.~\ref{fig:evolution}(A) shows the evolutionary tracks of magnetars in the $P-\dot{P}$ diagram connecting present to the birth stage. 
The corresponding death lines are also shown here, and the sources remain above their death line for the assigned $\chi_0$. Here, the birth stage corresponds to the termination point, $t=-t_{z}$, at which $\chi\rightarrow \pi/2$. For all the cases, we have taken a numerical precision up to $\sin^{2}\chi=0.9999$ at $t=-t_{z}$. The age measurement remains unaffected even for better precision since $t$ is asymptotic in the limit $\chi\rightarrow \pi/2$, as shown in Fig.~\ref{fig:evolution}(C). Note that the model can not predict the spin period at birth as it strongly depends on the numerical precision in the limit of $\chi\rightarrow \pi/2$.
The evolutionary tracks bend downwards after the initial stage from birth due to the magnetic field dissipation. Therefore, all the tracks naturally reach their asymptotic period, $P$. The time evolution of $\chi$ plays a very significant role in achieving a smaller $\dot{P}$ value for older magnetars. Note that the GW torque hardly plays any role in the spin-down evolution. Fig.~\ref{fig:evolution}(B) shows the magnetic field evolutions for magnetars during their life span.

\section{Conclusions} \label{sec:conclusions}

We solve self-consistently the spin-down evolutions of magnetars by considering a plasma-filled oblique rotator model with state-of-the-art magnetic field decay mechanisms. The presence of plasma affects the magnetospheric properties, and also acceleration gaps are formed in the vicinity of the star. The rotational period of magnetars increases over time due to the extraction of angular momentum by GW radiations, magnetic dipole radiations, and particle winds. These torques also change the obliquity angle $\chi$. Our principal conclusions are as follows: 

1. We unify all the SGRs and AXPs irrespective of their high as well as low $B_{c}$. In the peculiar case of SGR 0418+5729, we find a dipolar $B$-field of $1.0\times10^{14}$ G, consistent with the surface $B$-field inferred from the X-ray spectral analysis \citep{2011MNRAS.418.2773G} and the presence of a variable absorption feature \citep{2013Natur.500..312T} independently.

2. Unlike ordinary radio pulsars, the spin-down age, $\tau_{c}$, for magnetars is not a reliable indicator for their actual ages. We provide a novel approach to estimate the true age of a magnetar. For SGR 0418+5729, we find a realistic age of $\sim18$ kyr for the first time. The independent age measurement will confirm the genuine association between SGRs/AXPs and SNRs, hence the magnetar's birth site.

\acknowledgments

We thank Banibrata Mukhopadhyay for useful discussions and comments that helped to improve this work notably. We are grateful to the anonymous referee for stimulating comments that improved the manuscript.





\end{document}